\documentclass[preprint,showpacs,preprintnumbers,amsmath,amssymb,natbib]{revtex4}

\usepackage{graphicx}
\usepackage{epsfig}
\usepackage{dcolumn}
\usepackage{bm}

\def\be{\begin{equation}}
\def\ee{\end{equation}}
\def\ba{\begin{eqnarray}}
\def\ea{\end{eqnarray}}

\begin{document}

\title{Entanglement due to noncommutativity in the phase-space}

\author{Catarina Bastos\footnote{E-mail: catarina.bastos@ist.utl.pt}}

\affiliation{Instituto de Plasmas e Fus\~ao Nuclear, Instituto Superior T\'ecnico Avenida Rovisco Pais 1, 1049-001 Lisboa, Portugal}

\author{Alex E. Bernardini\footnote{On leave of absence from Departamento de F\'isica, Universidade
Federal de S\~ao Carlos, PO Box 676, 13565-905, S\~ao Carlos, SP, Brasil. E-mail: alexeb@ufscar.br}, Orfeu Bertolami\footnote{Also at Instituto de Plasmas e Fus\~ao Nuclear, Instituto Superior T\'ecnico,
Avenida Rovisco Pais 1, 1049-001 Lisboa, Portugal. E-mail: orfeu.bertolami@fc.up.pt}}

\affiliation{Departamento de F\'isica e Astronomia, Faculdade de Ci\^encias da Universidade do Porto, Rua do Campo Alegre, 687,4169-007 Porto, Portugal}

\author{{Nuno Costa Dias and Jo\~ao Nuno Prata}\footnote{Also at Grupo de F\'{\i}sica Matem\'atica, UL,
Avenida Prof. Gama Pinto 2, 1649-003, Lisboa, Portugal. E-mail: ncdias@meo.pt, joao.prata@mail.telepac.pt}}
\affiliation{Departamento de Matem\'{a}tica, Universidade Lus\'ofona de
Humanidades e Tecnologias Avenida Campo Grande, 376, 1749-024 Lisboa, Portugal}

\date{\today}

\begin{abstract}
The entanglement criterion for continuous variable systems and the conditions under which the uncertainty relations are fulfilled are generalized to the case of a noncommutative (NC) phase-space.
The quantum nature and the separability of NC two-mode Gaussian states are examined.
It is shown that the entanglement of Gaussian states may be exclusively induced by switching on the noncommutative deformation.
\end{abstract}

\maketitle

\paragraph*{Introduction -}

Besides its theoretical interest, the role of quantum entanglement has been acknowledged for its wide range of applications in quantum information protocols \cite{theory2} and quantum
communication \cite{theory1}.
On the theoretical side, one of the key results is the positive partial transposed (PPT) separability criterion \cite{Peres}, which provides a necessary and, in some cases, a sufficient condition for distinguishing between separable and entangled states in discrete quantum systems.
This criterion was extended to continuous variable systems \cite{Simon} by implementing the partial transpose operation as a mirror reflection in the Wigner phase-space.
An important application of the ``continuous'' PPT criterion is in the theory of quantum information of Gaussian states \cite{gauss01,Adesso}, which is at the core of the test beds for estimating quantum correlations \cite{Weedbrook}.
In this case, the PPT criterion yields both a necessary and sufficient separability condition
\cite{Adesso,Simon}.
As will be seen, Gaussian states are also quite useful for investigating the entanglement exclusively induced by a noncommutative (NC) deformation of phase-space.

Noncommutativity is believed to be an essential feature of quantum gravity and string theory \cite{Connes,Seiberg}.
In the context of quantum mechanics (QM) and quantum cosmology, phase-space noncommutativity has been shown to exhibit striking features with applications for the black hole singularity \cite{Bastos3} and the equivalence principle \cite{Bastos4}.
Furthermore, the quantum mechanical aspects of NC theories have focused on studies of the quantum Hall effect \cite{Prange,Belissard}, the gravitational quantum well for ultra-cold neutrons \cite{06A}, the Landau/$2D$-oscillator problem in the phase-space \cite{Nekrasov01,Rosenbaum}, and as a probe of quantum beating and missing information effects \cite{2013}.

The phase-space NC generalizations of QM are based on extensions
of the Heisenberg-Weyl algebra \cite{Bastos0,GossonAdhikari,08A}.
In this letter, the PPT entanglement criterion for continuous
variable systems is generalized to the case of a phase-space with
canonical momentum-momentum and position-position
noncommutativity. Two alternative formulations of the new
criterion are presented. The second formulation consists on a
condition on the smallest symplectic eigenvalue of the partially
transposed covariance matrix of the state, and is very efficient
from the computational point of view. We will use it to study in
detail the effect of noncommutativity on the quantum nature and
separability properties of Gaussian states. Most interesting is
the conclusion that noncommutativity alone can induce the
entanglement of Gaussian states. Besides its theoretical interest,
this result may become important for the experimental tests of NC
QM.

For the purpose of quantifying how noncommutativity affects the separability of quantum states, let us consider a bipartite quantum system described in terms of a $2n_A$-dimensional subsystem A (Alice) and a $2n_B$-dimensional subsystem B (Bob) with $n_A+n_B=n$.
One may write the collective degrees of freedom of the composite system
$\widehat{z} = (\widehat{z}^A,\,\widehat{z}^B)$, where $\widehat{z}^A =(\widehat{x}_1^A,
\cdots,\widehat{x}_{n_A}^A,\, \widehat{p}_1^A, \cdots, \widehat{p}_{n_A}^A)$ and $\widehat{z}^B
=(\widehat{x}_1^B, \cdots,\widehat{x}_{n_B}^B,\, \widehat{p}_1^B, \cdots, \widehat{p}_{n_B}^B)$ are the
corresponding generalized variables of the two subsystems, which obey the commutation relations
\begin{equation}
\left[\widehat{z}_i, \widehat{z}_j \right] = i \Omega_{ij}, \hspace{1 cm} i,j = 1, \cdots, 2n,
\label{eq1}
\end{equation}
where the associated matrix is given by ${\bf \Omega} = \left[\Omega_{ij} \right] \equiv Diag\left[\bf \Omega^A,\,\bf \Omega^B\right]$,
with ${\bf \Omega^K}$ a real skew-symmetric non-singular $2n_K \times 2 n_K$ matrix $(K=A,B)$ of the form
\begin{equation}
{\bf \Omega^K} = \left(
\begin{array}{c c}
{\bf \Theta^K} & {\bf I^K}\\
- {\bf I^K} & {\bf \Upsilon^K}
\end{array}
\right),
\label{eq2}
\end{equation}
where ${\bf \Theta^K} = \left[\theta_{ij}^K \right]$ and ${\bf \Upsilon^K} = \left[\eta_{ij}^K \right]$ measure the noncommutativity of the position-position and
momentum-momentum sectors of the subsystem $K = A,\,B$, ${\bf I^K}$ is the $n_K \times n_K$ identity matrix, and we have set $\hbar =1 $.

The NC structure can be formulated in terms of commuting variables by considering a linear Darboux transformation (DT), $\widehat{z} = {\bf S} \widehat{\xi}$, where $ \widehat{\xi}=( \widehat{\xi}^A,
\widehat{\xi}^B)$, with $ \widehat{\xi}^A= ( \widehat{q}_1^A, \cdots, \widehat{q}_{n_A}^A, \, \widehat{k}_1^A,
\cdots, \widehat{k}_{n_A}^A)$ and $ \widehat{\xi}^B= ( \widehat{q}_1^B, \cdots, \widehat{q}_{n_B}^B,\,
\widehat{k}_1^B, \cdots, \widehat{k}_{n_B}^B)$, satisfy the usual commutation relations:
\begin{equation}
\left[\widehat{\xi}_i, \widehat{\xi}_j \right] = i J_{ij} , \hspace{1 cm} i,j=1, \cdots, 2n.
\label{eq3}
\end{equation}
Here ${\bf J} = \left[J_{ij} \right]=Diag \left[{\bf J^A},{\bf J^B} \right] $ where
\begin{equation}
{\bf J^K} = - ({\bf J^K})^T = - ({\bf J^K})^{-1} = \left(
\begin{array}{c c}
{\bf 0} & {\bf I^K}\\
- {\bf I^K} & {\bf 0}
\end{array}
\right)~
\label{eq4}
\end{equation}
is the $2n_K \times 2n_K$ standard symplectic matrix for $K =
A,\,B$. The linear transformation ${\bf S} \in Gl(2n)$ is such
that ${\bf S} = Diag\left[{\bf S^A},\,{\bf S^B}\right]$, and ${\bf
\Omega} ={\bf S} {\bf J} {\bf S}^T  $, or equivalently ${\bf
\Omega^K} ={\bf S^K} {\bf J^K} ({\bf S^K})^T  $ for $K=A,B$.
Notice that the map ${\bf S}$ is not uniquely defined. If we
compose ${\bf S}$ with block-diagonal canonical transformations we
obtain an equally valid DT. We will see, however, that our main
results are independent of the particular choice of the map ${\bf
S}=Diag\left[{\bf S^A},\,{\bf S^B}\right]$.

\paragraph*{Noncommutative separability criterion -}

Let us assume that the composite quantum system is described by the density matrix $\rho$, function of the NC variables $\widehat{z}$.
The DT yields the density matrix $\widetilde{\rho} (\widehat{\xi}) = \rho \left({\bf S}
\widehat{\xi} \right)$, which is associated with the Wigner function: \small \be W \widetilde{\rho} (\xi) = \! {1\over{(2 \pi)^{n}}} \int_{\mathbb{R}^{n_A}}\!\! \!\!dy^A
\!\!\int_{\mathbb{R}^{n_B}} \!\!\!\!dy^B e^{- i \left(y^A \cdot k^A + y^B \cdot k^B\right)} \left\langle q^{A} +{y^A  \over 2},\,q^{B} +{y^B  \over 2} \left\vert\, \widetilde{\rho}
\,\right\vert q^{A} -{y^A  \over 2},\,q^{B} -{y^B  \over 2}\right\rangle . \label{eq5} \ee \normalsize Upon inversion of
the DT, we obtain the NC Wigner function
\cite{Bastos0}:
\begin{equation}
W \rho (z) = {1 \over \sqrt{\det {\bf \Omega}}} W \widetilde{\rho} ({\bf S}^{-1} z).
\label{eq6}
\end{equation}
If ${\bf \Sigma}$ denotes the covariance matrix of $W \rho$ and ${\bf \widetilde{\Sigma}}$ that of $W \widetilde{\rho}$, then the two are related by:
\begin{equation}
{\bf \Sigma} = {\bf S} {\bf \widetilde{\Sigma}} {\bf S}^T.
\label{eq7}
\end{equation}
A necessary condition for the phase-space function $W \widetilde{\rho}$ with covariance matrix ${\bf
\widetilde{\Sigma}}$ to be an admissible (commutative) Wigner function is that it satisfies the
Robertson-Schr\"odinger uncertainty principle (RSUP): ${\bf \widetilde{\Sigma}} + \frac{i}{2} {\bf J} \ge 0$.
That is, the matrix ${\bf \widetilde{\Sigma}} + \frac{i}{2} {\bf J}$ has to be a $2n \times 2n$ positive matrix
in $\mathbb{C}^{2n}$. From this condition and Eq.~(\ref{eq7}) we conclude that for $W \rho$ to be a
{\it bona fide} NC Wigner function, it has to satisfy the NC RSUP \cite{Bastos1}:
\begin{equation}
{\bf \Sigma} + {i \over 2} {\bf \Omega} \ge 0.
\label{eq8}
\end{equation}
For Gaussian states this condition is also sufficient \cite{Bastos1}.

A composite quantum system is separable if its density matrix
takes the form $\rho = \sum_{i=1}^{\infty} \lambda_i \rho_i^A
\otimes \rho_i^B$, where $0 \le \lambda_i \le 1$, for all $i \in
\mathbb{N}$, $\sum_{i=1}^{\infty} \lambda_i =1$ and ${\rho}_i^A$
(resp. ${\rho}_i^B$) is a density matrix which involves only
Alice's (resp. Bob's) coordinates $\widehat{z}^A$ (resp.
$\widehat{z}^B$). The associated Wigner function is:
\begin{equation}
W \rho (z) = \sum_{i=1}^{\infty} \lambda_i W \rho_i^A (z^A) W \rho_i^B (z^B).
\label{eq9}
\end{equation}
Using the DT we obtain the commutative counterpart in terms of the commutative variables
$\xi$:
\begin{equation}
W \widetilde{\rho} (\xi) = \sum_{i=1}^{\infty} \lambda_i W \widetilde{\rho}_i^A (\xi^A) W \widetilde{\rho}_i^B (\xi^B).
\label{eq10}
\end{equation}
To simplify the manipulation of the above results, let us define ${\bf \Lambda}$ to be the $2 n\times 2n$ matrix ${\bf \Lambda}= Diag\left[{\bf I^A},\, {\bf \Lambda^B}\right]$, with ${\bf \Lambda^B}= Diag\left[{\bf I},\, -{\bf I}\right]$.
Thus, the transformation
$\xi \mapsto {\bf \Lambda} \xi$
amounts to a mirror reflection of Bob's momenta.

According to the PPT criterion, if a Wigner function $W \widetilde{\rho} (\xi)$ is that of a separable state, then under the transformation
\begin{equation}
W \widetilde{\rho} (\xi) \mapsto W \widetilde{\rho}^{\prime} (\xi) = W \widetilde{\rho} ({\bf \Lambda} \xi),
\label{eq11}
\end{equation}
one obtains an equally admissible Wigner function.
Hence, if the state $\widetilde\rho$ is separable then
\begin{equation}
\widetilde{\bf \Sigma}^{\prime} + {i\over 2}{\bf J}\geq 0
\label{eq12}
\end{equation}
where $\widetilde{\bf \Sigma^{\prime}}$ is the covariance matrix
of $W \widetilde{\rho}^{\prime} (\xi)$. The transformation, Eq.
(\ref{eq11}), can be rewritten in terms of the NC variables as
follows:
\begin{equation}
W {\rho} (z) \mapsto W \rho^{\prime} (z) = W \rho ( {\bf D} z),
\label{eq13}
\end{equation}
where $W \rho (z) = {1 \over \sqrt{\det {\bf \Omega}}} W
\widetilde{\rho} ({\bf S}^{-1} z)$ and $W \rho^\prime (z) = {1
\over \sqrt{\det {\bf \Omega}}} W \widetilde{\rho}^\prime ({\bf
S}^{-1} z)$ are defined accordingly to Eq.~(\ref{eq6}) and ${\bf
D}= {\bf D}^{-1}= {\bf S}{\bf \Lambda}{\bf S}^{-1} = Diag[{\bf
I^A},\,{\bf S^B} {\bf \Lambda^B} ({\bf S^B})^{-1}]$. It follows
from Eq.~(\ref{eq7}) that the covariance matrices ${\bf
\Sigma^{\prime}}$ and ${\bf \widetilde\Sigma^{\prime}}$ of $W
\rho^\prime$ and $W \widetilde\rho^\prime$ are related by ${\bf
\Sigma^{\prime}} = {\bf S}{\bf \widetilde\Sigma^{\prime}} {\bf
S}^T$. Hence, the separability condition Eq.~(\ref{eq12}) can then
be written exclusively in terms of the NC objects
\begin{equation}
{\bf \Sigma^{\prime}} + {i\over 2}{\bf \Omega}\geq 0~. \label{eq13A}
\end{equation}
In addition, notice that ${\bf \Sigma^{\prime}} = {\bf D}{\bf \Sigma} {\bf D}^T$ where ${\bf \Sigma}$ is the covariance matrix of $W\rho$.
Defining ${\bf \Omega^{\prime}} = {\bf D}^{-1} {\bf\Omega} ( {\bf D}^T)^{-1}$ we obtain
\begin{equation}
{\bf \Sigma} + {i \over 2}   {\bf \Omega^{\prime}}  \ge 0~. \label{eq15}
\end{equation}
We point out that the matrix ${\bf \Omega^{\prime}}$ is simply given by
\begin{equation}
{\bf \Omega}^{\prime} = Diag\left[{\bf
\Omega^A},\,-{\bf \Omega^B}\right],
\label{eq15.1}
\end{equation}
where one has used the definition of ${\bf \Omega}$ and the fact
that ${\bf \Lambda^B}$ is an anti-symplectic transformation, i. e.
${\bf \Lambda^B} {\bf J^B} {\bf \Lambda^B} = - {\bf J^B}$. By
itself, this result is an elegant generalization of the result
from Ref.~\cite{Simon} stating that ${\bf J}$ is replaced by ${\bf
J}^{\prime} = Diag \left[{\bf J}^A, - {\bf J}^B \right]$ under
PPT. We stress the fact that Eq.~(\ref{eq15.1}) is valid assuming
that the DTs take the block-diagonal form ${\bf S}=Diag \left[{\bf
S^A},{\bf S^B}\right]$.

Finally, for Gaussian functions our results yield the following characterization:
A real normalizable phase-space Gaussian function $F(z)$ with covariance matrix ${\bf \Sigma}$ is the NC Wigner function of a quantum separable state if and only if ${\bf \Sigma}$ satisfies the Eqs.~(\ref{eq8}) and (\ref{eq15}) with respect to quantum nature and separability, respectively.

\paragraph*{Noncommutative symplectic spectrum and DT invariance -}

To proceed we reexpress the NC quantumness and separability
criteria in terms of the so-called NC symplectic spectrum. First
recall that the {\it symplectic spectrum} of a $2n \times 2n$ real
symmetric positive-definite matrix ${\bf \widetilde{\Sigma}}$ is
the set of eigenvalues of the matrix $2 i {\bf J}^{-1} {\bf
\widetilde{\Sigma}}$ and these eigenvalues are called the
Williamson invariants of ${\bf \widetilde{\Sigma}}$. They are all
positive and we denote by $\widetilde{\nu}_-$ the smallest
Williamson invariant. By Williamson's Theorem \cite{Simon}, one
can show that ${\bf \widetilde{\Sigma}}$ complies with the RSUP
(${\bf \widetilde{\Sigma}} + \frac{i}{2} {\bf J} \ge 0$) if and
only if $\widetilde{\nu}_- \ge 1$. By the same token, we call the
set of eigenvalues of $2i {\bf \Omega}^{-1} {\bf \Sigma}$ the {\it
NC symplectic spectrum} of ${\bf \Sigma}$ and the eigenvalues are
called the NC Williamson invariants. We denote by $\nu_-$ the
smallest NC Williamson invariant of ${\bf \Sigma}$.

Now consider a covariance matrix ${\bf \Sigma}$ and let ${\bf S}$ be any DT.
Suppose that ${\bf \widetilde{ \Sigma}}$ is related to ${\bf \Sigma}$ by Eq.~(\ref{eq7}).
Then we claim that the NC symplectic spectrum of ${\bf \Sigma}$ coincides with the symplectic spectrum of ${\bf \widetilde{ \Sigma}}$.
Indeed:
\begin{equation}
0 = \det(2i {\bf \Omega}^{-1} {\bf \Sigma} - \lambda {\bf I})= \det(2i ({\bf S} {\bf J} {\bf S}^T)^{-1} {\bf S} {\bf \widetilde{\Sigma}} {\bf S}^T - \lambda {\bf I}) = \det(2i {\bf J}^{-1} {\bf \widetilde{\Sigma}} - \lambda {\bf I}).
\label{eq16}
\end{equation}
We then have the following sequence of equivalent statements:
\begin{equation}
{\bf \Sigma} + \frac{i}{2} {\bf \Omega} \ge 0 \Leftrightarrow {\bf \widetilde{\Sigma}} + \frac{i}{2} {\bf J} \ge 0
\Leftrightarrow \widetilde{\nu}_- \ge 1 \Leftrightarrow \nu_- \ge 1.
\label{eq17}
\end{equation}
In other words, ${\bf \Sigma}$ satisfies the NC RSUP if and only if its smallest NC Williamson invariant satisfies $\nu_- \ge 1$.

Now we turn to the separability criterion Eq.~(\ref{eq15}).
Let $\nu_-^{\prime}$ denote the smallest NC Williamson invariant of ${\bf \Sigma^{\prime}} = {\bf D} {\bf \Sigma} {\bf D}^T $.
If ${\bf \Sigma}$ is the covariance matrix associated with a NC separable state, then $\nu_-^{\prime} \ge 1$.
To summarize, a necessary condition for ${\bf \Sigma}$ to be the covariance matrix of a NC quantum state is given by $\nu_- \ge 1$.
Likewise, $\nu_-^{\prime} \ge 1$ is a necessary condition for ${\bf \Sigma}$ to be the covariance matrix of a separable NC quantum state.
Again, if the state is a Gaussian both are sufficient criteria.

The NC symplectic spectrum of ${\bf \Sigma}$ depends only on ${\bf \Sigma}$ and on ${\bf \Omega}$ and hence it is manifestly independent of the DT ${\bf S}$.
Consequently, the NC quantum nature condition, $\nu_- \ge 1$, is also independent of ${\bf S}$.
On the other hand, the NC symplectic spectrum of ${\bf \Sigma^{\prime}}$ is also independent of ${\bf S}$, provided it is block-diagonal.
This is easily seen, if we consider the fact that the spectrum of $2i {\bf \Omega}^{-1} {\bf \Sigma^{\prime}}$ is the same as that of $2i {\bf \Omega^{\prime}}^{-1} {\bf \Sigma}$ and ${\bf \Omega^{\prime}}$ takes the form (\ref{eq15.1}) for all block-diagonal DTs.

\begin{figure}
\begin{center}
\includegraphics[scale=0.5]{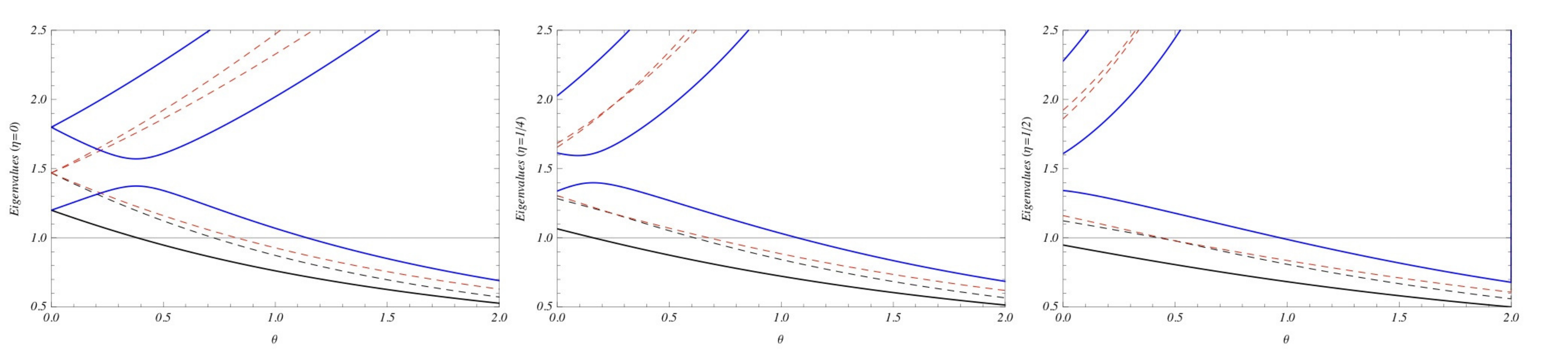}
\caption{Eigenvalues $\nu$ (dashed lines) and $\nu^\prime$ (solid lines) for $\theta = 0,\,1/4$,
and $1/2$ as function of $\eta$ in the range $[0,\,2]$. Black lines correspond to the respective smallest eigenvalues, $\nu_-$ and $\nu_-^\prime$. Notice that entanglement ($\nu_-^\prime < 1$) coexists with quantum
behavior ($\nu_- \geq 1$) for $\eta \neq 0$.} \label{fig:results1}
\end{center}
\end{figure}

\begin{figure}
\begin{center}
\includegraphics[scale = 0.7]{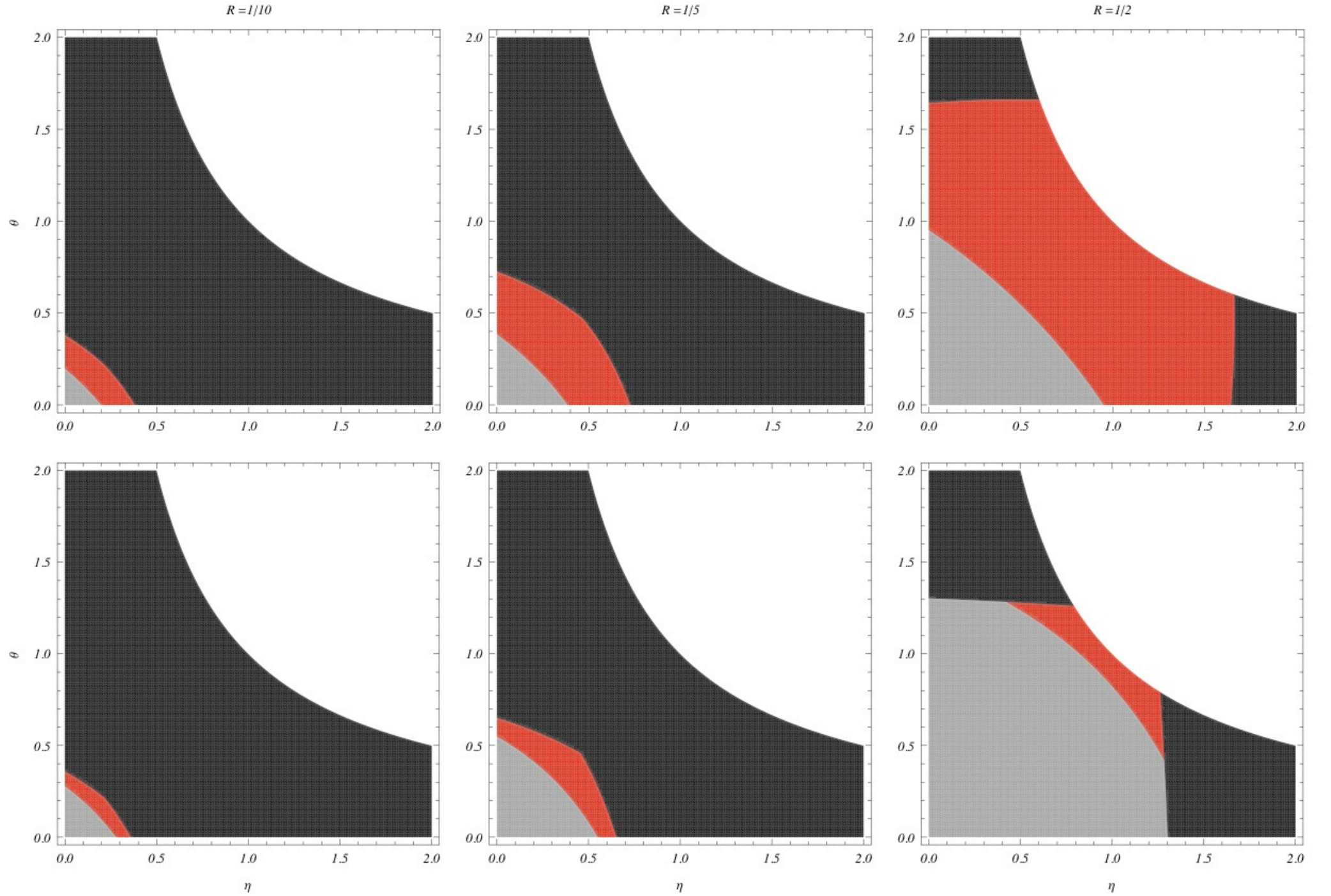}
\caption{The entanglement properties of the NC Gaussian states with the covariance matrix constrained by
$R=1/10,\, 1/5$, and $1/2$. Plots are for $n = R/3$ and $m = \sqrt{2}R/3$ (first line) and for $n = \sqrt{2}R/3$
and $m = R/3$ (second line). Gray and red regions correspond respectively to separable ($\widetilde{\nu}_- \geq 1$)
and entangled ($\nu_-^{\prime} < 1$) quantum states ($\nu_-^{\prime} \geq 1$). Black region denotes the violation of the
RSUP ($\nu_- < 1$) and the white region is out of the region bound by $\theta \eta \leq 1$.}
\label{fig:results2}
\end{center}
\end{figure}

\paragraph*{$2$-dim Gaussian state -}
In order to evince the entanglement induced by the noncommutative deformation, let us consider a Gaussian state,
\begin{equation}
F(z) = \frac{1}{\pi^4 \sqrt{\det {\bf \Sigma}}} \exp (- z^T {\bf \Sigma}^{-1} z),
\label{eq19}
\end{equation}
living on a $8$-dimensional NC phase-space with $n_A = n_B = 2$, ${\bf \Omega^A}= {\bf \Omega^B}$, $\theta_{ij} = \theta\epsilon_{ij}$ and $\eta_{ij} = \eta\epsilon_{ij}$, with $i,\,j = 1,\,2$, and where the covariance matrix is given by,
\small\be
{\bf \Sigma}={b\over 2}\left(
\begin{array}{c c}
{\bf I}_{4}&{\bf \gamma^T}\\
{\bf \gamma}&{\bf I}_{4}
\end{array}\right),~
\mbox{with}~
{\bf \gamma}=\left(
\begin{array}{c c c c}
n &  0 &  m & 0\\
0 &  n &  0 & -m\\
m &  0 & -n & 0\\
0 & -m &  0 & -n
\end{array}\right),
\label{eq20}
\ee
\normalsize
where $b,\,m$ and $n$ are real parameters which we assume, for simplicity, to be constrained by the relations $R = \sqrt{m^2 + n^2}$, $b = (1 + R)/(1 - R)$.
In this case, the DT corresponds to the map ${\bf S}= Diag[{\bf S^A},\,{\bf S^B}]$, with
\be
{\bf S^A}= {\bf S^B} = \left(
\begin{array}{c c c c}
\lambda &  0 &  0 & -\theta/2\lambda\\
0 &  \lambda &  \theta/2\lambda & 0\\
0 &  \eta/2\mu & \mu & 0\\
\eta/2\mu& 0 &  0 & \mu
\end{array}\right),
\label{eq21}
\ee
subject to $\lambda\mu = (1 + \sqrt{1 - \eta\theta})/2$, which ensures the map invertibility.
The NC parameters, $\theta$ and $\eta$, are real positive constants satisfying the condition $\theta\eta < 1$.
The smallest NC Williamson invariants of ${\bf \Sigma}$ and ${\bf \Sigma^{\prime}}$ are given by
\be
\nu_- ={1 \over {1-\eta\theta}}{{1+R} \over {1-R}}\sqrt{{{\omega_-} \over 2} -\sqrt{{{\omega_-^2} \over 4} -  \left(1-R^2\right)^2 (1-\eta  \theta )^2}},\nonumber
\ee
\be
\nu_-^{\prime} ={1 \over {1-\eta\theta}}{{1+R} \over {1-R}}\sqrt{{{\omega_+} \over 2} -\sqrt{{{\omega_+^2} \over 4} -  \left(1-R^2\right)^2 (1-\eta  \theta )^2}},\nonumber
\ee
respectively, where
$\omega_{\pm} = 2\left(1\pm n^2\right)+\left(1 \mp n^2\right) \left(\eta^2 + \theta^2\right) \pm 2 m^2 (1 + \eta\theta) + n(1 \mp 1)\vert\eta^2 - \theta^2\vert + 2m(1 \pm 1)(\eta + \theta)$,
which allows for examining the role of the NC parameters on the Gaussian entanglement.
The NC quantum nature and the separability of Gaussian states are ensured for $\nu_- \geq 1$ and $\nu_-^{\prime} \geq 1$, respectively.
Entangled quantum states are found in the range  $\nu_- \geq 1 >  \nu_-^{\prime}$ (cf. Fig.~\ref{fig:results2}).
Setting $\theta = 0$, the obtained NC Williamson invariants, $\nu_-$ and $\nu_-^{\prime}$, depend exclusively on $\eta$, as depicted in Fig. \ref{fig:results1}.
The standard (commutative) QM limit is obtained by setting $\theta  = \eta = 0$, which implies that $\nu_- = (1 + R)^{3/2}/(1-R)^{1/2}$ and $\nu_-^{\prime} = 1 + R$.
This means that for $0 \le R < 1$, all states are quantum  ($\nu_-\geq 1$), and separable ($\nu_-^{\prime} \geq 1$).
Fig.~\ref{fig:results2} shows how the NC phase-space ($\theta\neq 0 \neq \eta$) induces the entanglement of the corresponding Gaussian states.
The influence of the NC parameters, $\theta$ and $\eta$, on the quantum nature, separability and entanglement of states can be depicted for several values of $R$ (with a degeneracy associated to $m \leftrightarrow n$).

\paragraph*{Conclusions -}
In this work the PPT criterion for entanglement-separability was
extended to NC QM in phase-space, and formulated exclusively in
terms of the NC framework. Moreover, we studied the impact of
changing the NC structure on the quantum nature and separability
properties of Gaussian states and showed that the entanglement can
be exclusively induced by the NC structure of phase space.
Finally, we point out that these results may become increasingly
important in view of the recently observed violation of the
uncertainty principle \cite{RSUP}. 

{\em Acknowledgements -} The
work of CB is supported by the FCT (Portugal) grant
SFRH/BPD/62861/2009. The AEB work is supported by the FAPESP
(Brazil) grant 2012/03561-0. The work of OB is partially supported
by the FCT project PTDC/FIS/111362/2009. NCD and JNP have been
supported by the FCT grant PTDC/MAT/099880/2008.

\pagebreak

\end{document}